\documentstyle[11pt,baaa-eng,twoside,epsf,psfig]{article}
\begin{document}
\vskip 1.0cm
\markboth{J. P\'erez, P. Tissera, D.G. Lambas \& C. Scannapieco}{Metallicity and colours in galaxy pairs.}
\pagestyle{myheadings}


\vspace*{0.5cm}
\parindent 0pt{ COMUNICACI\'ON DE TRABAJO -- CONTRIBUTED  PAPER } 
\vskip 0.3cm
\title{Metallicity and colours in galaxy pairs in chemical hydrodynamical
simulations.}

\author{Josefa P\'erez}
\affil{Facultad de Ciencias Astron\'omicas y Geof\'\i sicas, UNLP, Argentina,
jperez@fcaglp.unlp.edu.ar}

\author{Patricia Tissera}
\affil{Instituto de Astronom\'\i a y F\'\i sica del Espacio, 
Buenos Aires, Argentina, patricia@iafe.uba.ar}

\author{Diego Garcia Lambas}
\affil{Observatorio Astron\'omico de la Universidad Nacional de C\'ordoba, 
Argentina, dgl@oac.uncor.edu}

\author{Cecilia Scannapieco}
\affil{Instituto de Astronom\'\i a y F\'\i sica del Espacio, 
Buenos Aires, Argentina, cecilia@iafe.uba.ar }

\begin{abstract}
Using chemical hydrodynamical simulations consistent with a $\Lambda$-CDM model,
we study the role played by mergers and interactions in the regulation of
 the star formation activity, colours and the chemical properties of galaxies in pairs. 
A statistical analysis of the orbital parameters in galaxy pairs ($r <100$ kpc $h^{-1}$) shows that the star formation (SF) activity correlates strongly with
the relative separation and weakly with the relative velocity,
indicating that close encounters ($r <30$ kpc $h^{-1}$) can increase
the SF activity to levels higher than that exhibit in galaxies without a close companion. Analysing the internal properties of interacting systems, we find that their stability properties  also play a role in the regulation the SF activity (P\'erez et al 2005a). Particularly, we find that 
the passive star forming galaxies in pairs are statistically 
more stable with deeper potential wells
and less leftover gas than active star forming pairs.
In order to compare our results with observations, we also build a 
projected catalog of galaxy pairs (2D-GP: $r_{p} <100$ kpc $h^{-1}$ and 
$\Delta V <350$ km $s^{-1}$), constructed by projecting the 3D sample in different random directions. In good agreement with  observations (Lambas et al 2003), our results indicate that galaxies with $r_{p} < 25$ kpc $h^{-1}$ (close pairs) show
an enhancement of the SF activity with respect to galaxies without
 a close companion. 
All the properties studied for galaxy pairs are analysed in
the 2D and 3D simulated catalogs, allowing us to 
assess the contamination level introduced by spurious pairs.
Consistently with observational estimations (Nikolic et al 2004), we
find that the percentage of spurious pairs decreases with the relative
separation, representing almost a $30\%$ for the 2D-GP and a $19\%$ for close pairs.
We also analyse the environmental effect on the star formation (SF) 
activity for both, pairs and isolated galaxy samples, finding the expected SFR-local density relation (Gomez et al 2003), with a significantly stronger dependence for close pairs.
Finally, we analyse the colour and chemical properties of galaxies in pairs
in order to investigate the effect of interactions on 
the bimodal colour distribution  observed in galaxies 
(Balogh et al. 2004; Tissera et al. 2005) and the mass-metallicity relation
for the stellar population (Tremonti et al. 2004).

\end{abstract}

\begin{resumen} 
Usando simulaciones 
qu\'\i micas  hidrodin\'amicas consistentes con un modelo $\Lambda$-CDM,
estudiamos el rol de las interacciones y colisiones de galaxias sobre
la actividad de formaci\'on estelar (SF), los colores y las propiedades 
qu\'\i micas de las mismas.
El an\'alisis estad\'\i stico de los efectos de las interacciones
entre galaxias sobre la actividad de formaci\'on estelar
como funci\'on de los par\'ametros orbitales, muestra que los
encuentros cercanos ($r <30$ kpc $h^{-1}$) pueden inducir una
 actividad de formaci\'on estelar superior a la encontrada en sistemas
aislados. Sin embargo, encontramos que la estabilidad de los sistemas
gal\'acticos tambi\'en juega un rol fundamental en la regulaci\'on de
este proceso (P\'erez et al 2005). As\'\i , las galaxias en pares con baja actividad de SF
tienden a ser m\'as estables (con pozos de potencial m\'as profundos)
que las galaxias en pares con alta taza de formaci\'on estelar.
A fin de comparar nuestros resultados con los observacionales,
construimos un cat\'alogo proyectado de pares de galaxias 
(2D-GP), siguiendo los criterios de selecci\'on utilizados por  
Lambas et al (2003): $r_{p} <100$ kpc $h^{-1}$ and $\Delta V <350$ km $s^{-1}$. 
Consistentemente con las observaciones, encontramos 
que los pares cercanos ($r_{p} < 25$ kpc $h^{-1}$) muestran niveles
de SF superiores a aquellos correspondientes a galaxias aisladas.
Comparando los cat\'alogos 2D-GP y 3D-GP,
estimamos los posibles efectos de proyecci\'on sobre
los resultados observacionales, analizando la contribuci\'on
de los pares espurios en las simulaciones. De acuerdo con estimaciones
observacionales (Nikolic et al 2004), nuestros resultados muestran que el porcentaje de pares espurios disminuye con la distancia relativa, encontrando un
$30\%$ para 2D-GP y un $19\%$ para los pares cercanos.
Por otro lado, analizamos los efectos ambientales sobre la actividad
de formaci\'on estelar, tanto para las galaxias en pares como para aquellas 
sin un vecino cercano. Encotramos, en buen acuerdo con las observaciones (Gomez 
et al 2003), que existe una relaci\'on entre la SF y la densidad local para 
ambas muestras con una mayor dependencia para los pares cercanos.
Finalmente, investigamos los efectos que producen las interacciones de galaxias sobre el color y las propiedades qu\'\i micas de las mismas, con el fin 
de evaluar su rol en la determinaci\'on de la
distribuci\'on bimodal de colores observada 
(Balogh et al. 2004; Tissera et al. 2005) y la relaci\'on 
masa-metalicidad de la poblaci\'on estelar  (Tremonti et al. 2004). 

\end{resumen}

\section{Introduction}

Observations show that mergers and interactions can  induce star
formation (SF) activity  in galaxies (e.g. Larson and Tinsley 1978).
Barton et al. (2000) and Lambas et al. (2003, LTAC03) analysed a sample
of pair galaxies in the field finding a clear correlation between  the proximity 
in projected distance and radial velocity of two galaxies and their SF activity.
On the other hand, numerical simulations of pre-prepared mergers showed that
interactions between axisymmetrical systems without bulges might induce gas 
inflows to the central region of the systems, triggering starburst episodes
(Mihos  and Hernquist 1996).
Results from the study of the effects of mergers in
the SF history of galactic objects in  cosmological hydrodynamical simulations 
(Tissera et al. 2002), indicate
that, during some mergers events, gaseous disks could experience
two starbursts depending on  the characteristic of the potential well.

Recently, observational results obtained by Balogh et al (2004) confirm a bimodal colour distribution, which segregates galaxies into blue and red populations. They find that this bimodality is well fitted using two Gaussian distributions. Analysing this colour distributions for an important range of magnitudes and densities, they find that while the characteristics of the
Gaussians depends strongly with luminosity, seems not to change with the local density. According to Balogh et al (2004) this invariability with the environment suggest that the process of transforming blue into red galaxies has
to be very efficient and to overcome the effects of environment.

In this work, we will investigate how effective interactions and mergers of galaxies are in the regulation of SF activity and if they are responsible in determining the other internal properties of galaxies like colours and chemical abundances. If the Universe is consistent with a hierarchical scenario, then, interactions and mergers are of utmost importance to understand the effects and efficiency that these physical processes have on the life of galaxies.
Thus, we will focus on a statistical analysis  of galaxy pairs  in a
hierarchical scenario  with the aim at confronting this scenario with
recent observational results of galaxy pairs. Details can be found in
P\'erez et al. (2005a, 2005b).

\section {Results}

We have analysed 
a $\Lambda$-CDM simulation
($\Lambda=0.7$, $\Omega=0.3$, $H=100 h$ ${\rm km s^{-1} Mpc^{-1}}$ with $ h=0.7$) run with  the chemical cosmological GADGET-2 (Scannapieco
et al. 2005). 
From this simulation we constructed the  2D-GP and 3D-GP catalogs.
In order to unveil the effects of interactions,
we also build 
 the respective galaxy control samples defined  
by galaxies without a close companion within the corresponding thresholds
in relative separation and velocity used to define each GP catalogs (P\'erez et al 2005a).
For each simulated galaxy, we estimated the stellar birthrate parameter
$b$, defined as the present level of SF activity of a galaxy
normalized to its mean past SF rate and the absolute magnitudes in different wavelenghts (De Rossi et al 2005).

The analysis of the SF activity for galaxies in the 3D-GP catalog as a function
of their orbital parameters shows that close encounters (pairs with a relative
separation less $30\pm 10$ kpc $h^{-1}$) can enhance the SF activity at higher
levels than those measured for galactic systems without a close companion.
On the other hand, the SF activity seems to correlate more weakly with the relative velocities. However, we also find almost a $50\%$ of passive star forming galaxies in close pairs, suggesting that the internal properties of these systems are also
present in the SF process. Effectively, we find that the triggering of SF by tidal interactions is also regulated by the stability of the systems and the gas reservoir. The analysis of the passive SF close pairs show that part of these
systems have experienced recent star formation activity and the rest shows deeper potential well and are less leftover gas than galaxies in pairs with strong SF activity. 

When the projected catalog of pair is analysed, the same global trends detected in the 3D-GP sample are found, although a shrinking in the enhancement threshold in projected distance is observed. It value 
drops with respect to that found in 3D to  $\sim 25 \pm 5$ kpc $h^{-1}$ in good agreement with observational results (LTAG03).
This shrinking in the threshold is produced by both geometrical projection 
effects and spurious pairs. In order to separate these both effects, we have removed spurious pairs from the  2D-GP sample by checking their 3D relative separations. Consistently with previous works (Alonso et al. 2004), we found that $30\%$ of the pairs  in  2D-GP sample
are spurious. This percentage reduces to    $19\%$ for  2D close pairs
($r_{p}<25 $ kpc $h^{-1}$ and $\Delta cz<100 $ km s$^{-1}$).

Analysing the dependence of the SF on environment with the projected local density parameter (defined like in the observational analysis as $\Sigma=6/(\pi d^{2}_{6})$, with $d_{6}$ the projected distance to the $6^{th}$ 
neighbor brighter than $M_{r}=-20.5$), we  detect the expected SFR-local 
density relation (Gomez et al 2003) 
for both galaxies in pairs and without a near companion, with a stronger dependence for close pairs. The important decrease in the SF activity and the significant increase of the fraction of passive SF members from low to high
density regions for galaxies in close pairs suggest that interactions might a relevant role in the origin of the SF-density relation.

In order to infer the effect of interactions on the colour distribution of galaxies, we also evaluate the $u-r$ colour for the simulated galaxies in pairs  compared with that found for galaxies without a near companion.
According results obtained by Balogh et al (2004), we adopt the value $u-r=1.8$  
to segregate red and blue galaxies. We found that the mean colours of the blue
and red peaks for pairs are at $<u-r>\approx 1.60$ and $<u-r>\approx 2.06$, with
similar values for the control sample. However, while pairs exhibit a clear bimodal colour distribution with a 58 per cent of galaxies in the blue peak, the
control sample is more consistent with an unimodal distribution with an excess of red systems (26 per cent of galaxies in the blue peak). Comparing 
the simulated control sample with observations (Balogh et al 2004), we also find an excess of the fraction of red galaxies. 
Part of this red excess might be produced by the high efficiency in the transformation of gas into stars of our simulations which is not regulated by supernova energy feedback. However, tidal torques generated by interactions can compress  the leftover gas  in a short-time producing strarbursts which might be the responsible of producing the bimodal distribution found for galaxies in pairs.
To gain  insight in this analysis, we divide the galaxy pair catalog into merging ($r<30 $ kpc $h^{-1}$) and interacting (30 kpc $h^{-1} <r<25 $ kpc $h^{-1}$) pairs. Although the bimodal distribution is present in both subsamples, the
fraction of blue galaxies is higher than the red one for the merging systems,
with the opposite result for the interacting pairs. This results is consistent with the present level of SF activity found for merging and interacting systems.
The former has a $36\%$ of active SF galaxies  (SF activity higher than for the control sample), while the latter has only a $10\%$ of active SF galaxies. 
The SF activity in the recent past for the currently passive SF systems is also
responsible for the colour distribution of galaxies. As shown in P\'erez et al (2005b), we find
that the fraction of currently passive SF systems which have experienced strong activity in the recent 0.5 Gyr (F$^{*}$) anticorrelates with the $u-r$ colour.
The contribution to the blue colours comes from galaxies that independently of their current SF activity, have experienced an strong SF activity in the recent last 
0.5 Gyr.

Finally, we analyse the chemical properties of the interstellar medium (ISM) and the stellar population (SP) of galaxies in pairs. While for SP exhibits a clear excess of their chemical abundance respect to that found for galaxies without a near companion, the ISM enriched by the new stars, has different levels of chemical contamination depending on the SF activity of the galaxy.
Particularly, if we segregate galaxies into active and passive according their
current SF level, we find that SP of passive SF galaxies are more enriched than the active ones. On the contrary, ISM of passive galaxies are less enriched  than the ISM of the active ones. Interesting, we also find that the distribution of the chemical abundance of the ISM of these systems has a similar behaviour with the relative separation between the members of the pair than that found for the SF
activity. In other words that the ISM of currently passive SF pairs shows an enhancement of their chemical enrichment respect to that measured in the control sample for very close systems ($r_{p}<25 $ kpc $h^{-1}$). The analysis of the recently past SF activity of these galaxy pairs
shows that, although these systems are currently passive SF ones, they have experienced strong SF activity in the recent past
which has contributed to enhance the chemical abundance in their ISM.

\section{Conclusions}

From the analysis of the 3D-simulated galaxy pair catalog, we conclude that close galaxy interactions ($r<30 $ kpc $h^{-1}$) can be correlated with  an enhancement of SF activity  at higher levels than those measured for galactic systems without a close companion.
We also found that the internal dynamical stability of galactic systems plays an 
important role as it can be deduced from the presence of an anticorrelation signal between the deepness of the potential well and the
star formation activity. 

The construction of a projected galaxy pair catalog allowed us, firstly, to make a suitable comparison with observations and then, to analyse the contamination effects introduced by spurious pairs. In a good agreement with observational results, we find that all trends observed for the 3D-GP catalog
are reproduced in a similar way for the 2D-GP one. On the other hand,
the analysis of the spurious pairs shows that the contamination level increases
with the relative separation.
Consistently with previous works (Alonso et al. 2004), we find that almost a $30\%$ of the pairs  in  2D-GP sample
are spurious.

The environmental effects on the interacting pairs were analysed, defining the local density using the projected distance to the $6^{th}$.
We find a dependence of the SF on local density which is consistent with the observed SF-density relation (Gomez et al 2003). It yields a decrease in the level of SF activity and an increase in the fraction of passive SF systems with increasing local density. Although, this relation is found for both pairs and
control samples, we find that it is significantly stronger for close pairs suggesting the fundamental role of interaction in driving the SF-density relation.

The analysis of the colours and chemical properties of galaxies also seem to
be regulated by the interactions. We find a clear bimodal colour distribution
for galaxies in pairs with a blue peak populated basically by galaxies with an strong, currently or recently past SF activity.
Finally, analysing the chemical properties of galaxy pairs, we find that while
SP seems to store information about the history of mergers and interactions, the chemical abundance of the ISM seems to be contaminated by the new stars, reflecting in this way the effects of the present interactions.
Our results are in very good agreement with recent observational works. Regarding the bimodal colour distribution, we 
found that
although interactions and merger are  able to explain the bimodal colour distribution, supernova energy feedback seems to be required
to prevent an excessive transformation  of gas into stars and to obtain the detail characteristics of this distribution.

\acknowledgments 
This work was partially supported by the Consejo Nacional
de Investigaciones Cient\'\i ficas y T\'ecnicas and Fundaci\'on Antorchas. 
Simulations
were run on Ingeld PC Cluster funded by Fundaci\'on Antorchas. We
thank the LOC of this meeting for their help made our participation 
possible. 
Patricia B. Tissera thanks the Aspen Center for Astrophysics for the hospitality during the Summer Workshop 2004.


\begin{references}
\reference Alonso, M. S., Tissera, P. B., Coldwell, G., Lambas D. G. 2004,MNRAS,
352,1081.
\reference Balogh M. L., Eke V., Miller C., et al., 2004, MNRAS, 348, 1355.
\reference Barton E. J., Geller M. J., Kenyon S. J., 2000, \apj, 530, 660.
\reference Lambas, D. G., Tissera, P. B., Alonso, M. S. Coldwell, G. 2003,
MNRAS, 346,1189 (LTAG03).
\reference Larson, R. B., Tinsley, B. M., 1978, \apj, 219, 46.
\reference Mihos, J. C., Hernquist, L., 1996,\apj,  464, 641.
\reference Navarro J.F. \& White S.D.M., 1994, MNRAS 267, 401.
\reference Nikolic B., Cullen H., Alexander P., 2004, MNRAS 355, 874.
\reference P\'erez, M.J., Tissera P.B., Lambas, D. G., Scannapieco, C. 2005, 
A\&A accepted (astro-ph/0510327)
\reference P\'erez, M.J., Tissera P.B., Lambas, D. G., Scannapieco, C. 2005,
\reference Scannapieco C., Tissera P.B., White S.D.M. \& Springel V., 2005, 
MNRAS  accepted (astro-ph/0505440)
\reference Tissera P.B, Dom\'{i}nguez-Teneiro R., Scannapieco C. \& S\'aiz A., 
2002,
MNRAS, 333, 327.
\end{references}
\end{document}